\documentclass[runningheads]{llncs}[10pt]
\usepackage{graphicx}
\usepackage[dvipsnames]{xcolor}
\usepackage{import}
\usepackage{bibnames}
\usepackage{hyperref}
\usepackage[linesnumbered,ruled,vlined]{algorithm2e}
\SetKwInput{KwInput}{Input}               
\SetKwInput{KwOutput}{Output}
\usepackage{xurl}
\usepackage{comment}
\usepackage{multirow}
\begin{document}
\title{MAD-EN: Microarchitectural Attack Detection through System-wide Energy Consumption}
\titlerunning{MAD-EN: Microarchitectural Attack Detection}
%
\author{Debopriya Roy Dipta \and
Berk Gulmezoglu}
\authorrunning{D. Dipta et al.}

%
\institute{Iowa State University,\\ Ames, IA 50010, USA}
%
\maketitle              
\vspace{-5mm}
\begin{abstract}

Microarchitectural attacks have become more threatening the hardware security than before with the increasing diversity of attacks such as Spectre and Meltdown. Vendor patches cannot keep up with the pace of the new threats, which makes the dynamic anomaly detection tools more evident than before. Unfortunately, previous studies utilize hardware performance counters that lead to high performance overhead and profile limited number of microarchitectural attacks due to the small number of counters that can be profiled concurrently. This yields those detection tools inefficient in real-world scenarios. 

In this study, we introduce \texttt{MAD-EN} dynamic detection tool that leverages system-wide energy consumption traces collected from a generic Intel RAPL tool to detect ongoing anomalies in a system. In our experiments, we show that CNN-based \texttt{MAD-EN} can detect 10 different microarchitectural attacks with a total of 15 variants with the highest F1 score of 0.999, which makes our tool the most generic attack detection tool so far. Moreover, individual attacks can be distinguished with a 98\% accuracy after an anomaly is detected in a system. We demonstrate that \texttt{MAD-EN} introduces 69.3\% less performance overhead compared to performance counter-based detection mechanisms.


\end{abstract}
\section{Introduction} \label{sec:introduction}

Microarchitectural attacks have demonstrated that fundamental chip design optimizations without considering security implications can lead to security and privacy issues on Intel, ARM, and AMD processors~\cite{zankl2021side}. These attacks affect billions of users on personal computers, cloud servers, and mobile phones through the exploitation of essential microarchitectural components~\cite{inci2016cache,lipp2018meltdown}. More importantly, these attacks can be performed from malicious third party applications or visited web pages in a browser without admin privileges, which tremendously increases the potential threat of microarchitectural attacks in the wild. 

Adversaries steadily discover new microarchitectural attacks to leak confidential information by leveraging microarchitectural design choices in commercial chips. As many processor designs were optimized to increase the performance with new fundamental features decades ago, it is not trivial to modify the microarchitectural designs in the new chip generations. Therefore, many existing chip designs are still subject to known microarchitectural attacks~\cite{barberis2022branch}. With the increasing diversity in microarchitectural attacks, cryptographic libraries such as OpenSSL have no further support to patch their source codes against microarchitectural attacks~\cite{openssl_nofix}. Hence, many attacks are still applicable on several cryptographic implementations until current fundamental microarchitecture design choices are changed entirely. 

One of the common threat detection techniques is to distinguish malicious and vulnerable code snippets through static analysis before they are executed. Tol et al.~\cite{tol2021fastspec} detect Spectre-vulnerable code snippets with Natural Language Processing (NLP) techniques without observing the dynamic behavior of a code segment. Similarly, Irazoqui et al.~\cite{irazoqui2018mascat} distinguish a subset of microarchitectural attacks from benign applications by identifying most commonly used instructions (e.g., rdtsc, lfence) in the attack codes. Unfortunately, several attacks~\cite{schwarz2017malware,shusterman2021prime+} showed that static features are not reliable to detect malicious attack codes as small perturbations and alternative instructions in a malicious code can be used to bypass the static detection tools.

The lack of efficient static tools encouraged both academia and industry to design dynamic attack detection techniques by monitoring a set of in-built sensors in modern processors. Since attacks generally target microarchitectural components, several studies focus on the utilization of hardware performance counters (HPCs) to collect low-level microarchitectural information during the attack execution to identify malicious behavior in a system~\cite{briongos2018cacheshield,chiappetta2016real,gulmezoglu2019fortuneteller}. Even high-end Intel processors deploy an anomaly detection mechanism based on performance counters through Intel Threat Detection Technology~\cite{intel_tdt} to detect malicious execution patterns. In a typical dynamic detection mechanism after a predetermined set of performance counters is monitored, the collected counter values are analyzed with several Machine Learning (ML) and Deep Learning (DL) algorithms through either supervised~\cite{chiappetta2016real} or unsupervised~\cite{briongos2018cacheshield,gulmezoglu2019fortuneteller} training. Even though the proposed techniques achieve higher detection rates for a limited number of attacks and targets, there are several disadvantages of using HPCs for dynamic detection tools: i) The number of concurrent performance counter sampling from Performance Monitoring Unit (PMU) is limited to three or four counters in current processors, which leads to limited coverage of microarchitectural attacks, ii) the performance overhead introduced by the counter monitoring tools is considerably high, creating higher power consumption for benign applications, iii) performance counters may not be consistent between the chip generations i.e. branch monitoring counters are deprecated starting from 12th generation Intel Core processors~\cite{intel_deprecate}.

Processors have many in-built sensors to investigate power consumption and temperature variations under several workloads. Intel provides an interface, namely running average power limit (RAPL)~\cite{david2010rapl}, to measure power consumption from core, DRAM, and offcore components. It has been shown that the RAPL interface enables adversaries to collect high resolution power consumption traces that can be used to leak secret keys~\cite{lipp2021platypus}. However, the interface has not been investigated for large-scale anomaly detection in commercial systems yet. In this study, we create a novel framework to detect microarchitectural attacks by analyzing power consumption of a processor through the RAPL interface in real time. We show that each attack has a specific fingerprint on power consumption traces that can be distinguished from benign applications with the help of Convolutional Neural Networks (CNNs). Since the power consumption traces provide more generic information system-wide compared to the performance counter values, we can detect in total 15 different microarchitectural attack variants with a single sensor. In summary, our contributions are as following:

\begin{itemize}
    \item We create an open-source energy consumption dataset for microarchitectural attacks and benign applications through RAPL interface \footnote{The dataset is available in GitHub: \url{https://github.com/Diptakuet/MAD-EN-Microarchitectural-Attack-Detection.git}}.
    \item The power traces are used to train a CNN model, and ongoing microarchitectural attacks can be detected with an F-score of 0.999.
    \item After an attack is detected, we can distinguish the correct attack type with 98\% success rate in 7.5 seconds.
    \item We show that our detection tool has considerably lower performance overhead compared to counter-based detection tools.
\end{itemize}

\noindent\textbf{Outline:} The rest of the paper is organized as follows: Section~\ref{sec:background} provides background on covered microarchitectural attacks and the Intel RAPL interface. Section~\ref{sec:related_work} gives an overview of dynamic detection techniques. Section~\ref{sec:prediction_model} introduces \texttt{MAD-EN} dynamic detection tool. Section~\ref{sec:evaluation} evaluates the applicability of \texttt{MAD-EN} tool in a real-world settings. Section~\ref{sec:discussion} discusses potential shortcomings of \texttt{MAD-EN}. Finally, Section~\ref{sec:conclusion} concludes the study.

\vspace{-4mm}
\section{Background}\label{sec:background}
\vspace{-2mm}
In this setion, we provide background on the targeted microarchitectural attacks and the Intel RAPL interface.
\vspace{-5mm}
\subsection{Micro-architectural Attacks} \label{subsec:Micro_attack}

\noindent\textbf{Flush+Reload (F+R)} exploits the page sharing feature between two running processes in the system~\cite{yarom2014flush+}. The entire attack consists of three principal stages--flush stage, victim access stage, and reload stage~\cite{gulmezouglu2015faster}. In the first stage, the attacker flushes the targeted memory block from the cache using the \textit{clflush} instruction. The second stage corresponds to a specific wait period to allow the victim access to the targeted memory line. At the final stage, the attacker reloads the memory block and measures the access time to detect victim activity. F+R attack has been utilized to recover secret keys from the victim's user-space~\cite{gulmezouglu2015faster}. 

\noindent\textbf{Flush+Flush (F+F)} exploits the execution time of the \textit{clflush} instruction~\cite{gruss2016flush+}. An attacker can distinguish whether a targeted memory line is cached or not based on the execution time of the \textit{clflush} instruction. If the data is cached, the execution time of the \textit{clflush} instruction will be comparatively higher since the instruction has to initiate eviction on all the local caches. F+F attack is considered as a stealthy attack as it does not require to make any memory accesses, thus keeping it undetectable based on cache misses and hits information. F+F attack can be utilized to launch AES T-table attacks or keystroke detection~\cite{gruss2016flush+}.

\noindent\textbf{Prime+Probe (P+P)} attack allows an attacker to monitor a specific cache set by filling all the ways with attacker-controlled memory blocks~\cite{liu2015last}. This attack is more generic than flush-based attacks as it does not require any page sharing and a shared cache level is sufficient to perform the P+P attack. It can be implemented in cross-core platforms to leak cryptographic keys in the cloud~\cite{inci2016cache}, web page fingerprinting~\cite{oren2015spy}, and user input detection~\cite{lipp2016armageddon}. 

\noindent\textbf{PortSmash}~\cite{aldaya2019port} attack targets the Simultaneous Multi-threading (SMT) architecture of the modern processors. 
PortSmash exploits timing information obtained from port contention in the execution units to recover secret keys. It has high adaptability with different hardware configurations, and may target several different ports to accommodate various scenarios, while keeping the prerequisites to its minimal.

\noindent\textbf{TLBleed} is a timing side-channel attack that can bypass the state-of-the-art protections against cache side-channel attacks. Gras et al. \cite{gras2018translation} exploits the SMT architecture that allows sharing resources between sibling threads while executing concurrent TLB accesses. In TLBleed, the attacker creates TLB eviction sets and monitors them to determine the TLB entries that are used for specific virtual memory translations by a victim. The attack has been proved to be capable of compromising EdDSA and RSA secret keys, even though the cache side-channel attack defenses were enabled \cite{gras2018translation}.

\noindent\textbf{Spectre} attacks~\cite{kocher2019spectre} take advantage of speculative execution feature in modern processors. Speculative execution allows processors to perform out-of-order execution speculatively to enhance the overall performance. In Spectre attacks, an attacker mistrains the branch prediction unit (BPU) to perform speculative execution in the wrong direction and leaks confidential data from the victim's user-space by encoding the secret into microarchitectural components. 

\noindent\textbf{Medusa} is a variant of microarchitectural data sampling (MDS) attacks~\cite{moghimi2020medusa}
, which only targets specific memory operations; thus making the attack more focused and effective compared to other MDS variants. Medusa is capable of leaking data from Write Combining (WC) memory operations. Moghimi et al.~\cite{moghimi2020medusa} demonstrated that Medusa can recover RSA keys, information from kernel data transfer, and so on.

\noindent\textbf{ZombieLoad} is a variation of Meltdown-type attacks, which exploits the fill-buffer that is shared by all logical threads in a core~\cite{schwarz2019zombieload}. During certain microarchitectural conditions, such as faulting load instruction that requires to be re-issued, the load may first read the stale values of the previous memory operations from either the current or sibling-thread before being re-issued. This feature enables attackers to perform transient execution attacks to leak information. 
Moreover, this attack is not limited by any privilege boundaries, and capable of leaking information even in Meltdown and MDS resistant processors.

\noindent\textbf{Fallout} is a Meltdown-type attack, in which an attacker leverages Write Transient Forwarding (WTF) to obtain unprivileged access to read kernel writes from user space. Additionally, it can also create a side-channel on the TLB, named Store-to-Leak, which is capable of breaking KASLR and ASLR from JavaScript~\cite{canella2019fallout}. 

\noindent\textbf{Branch History Injection (BHI)} is a new primitive of cross-privileged Branch Target Injection (BTI) or Spectre v2 attack. 
Barberis et al.~\cite{barberis2022branch} demonstrated that the BHI attack can bypass the newest hardware defenses--Intel eIBRS (Enhanced-Indirect Branch Restricted Speculation)~\cite{eIBRS} and Arm CSV2~\cite{grayson2020evolution}. BHI is still feasible since the Branch History Buffer (BHB) is not completely isolated among all privilege levels. Although eIBRS and CSV2 prevent any unprivileged users from injecting entries to the predictor for the kernel, the study showed that it is possible to exploit the dependency of the predictor on global history to force the kernel to miss-predict by manipulating the history from the user-space for leaking information.

\vspace{-5mm}
\subsection{Intel Running Average Power Limit (RAPL)} \label{subsec:intel_rapl}
\vspace{-1mm}

The power capping framework in Linux kernel provides a uniform \textit{sysfs} interface namely \textit{powercap} that allows to monitor and limit the power consumption of all the devices from the admin privileged space. In addition, the framework provides a common API for all the drivers whose settings are exposed to the privileged space. The \textit{powercap} interface includes multiple power capping drivers and tools, such as Intel Running Power Average Limit (RAPL), and Idle Injection. The Intel RAPL tool was first introduced for Intel Sandy Bridge architecture~\cite{khan2018rapl} that allows to monitor and control several energy/power consumption attributes from different power zones. The RAPL interface groups the entire processor in four power domains and allows to control/monitor power attributes in the individual domains. The available power domains of the RAPL interface are Package (pkg), Power Plane 0 (PP0), Power Plane 1 (PP1), and DRAM domain.

\noindent\textbf{Package Domain (Pkg)} allows to monitor/control accumulated energy consumption of the entire socket of a processor, which includes all the cores, graphics, and uncore components.

\noindent\textbf{Power Plane 0 (PP0)} incorporates all the cores in the socket. Thus, the energy consumption attribute within the PP0 domain can be utilized to measure energy consumption of the cored components of the processor.

\noindent\textbf{Power Plane 1 (PP1)} measures the energy consumption of the uncore components, such as GPUs, in the socket.

\noindent\textbf{DRAM} power domain provides access to monitor the energy consumption of the main memory. It is to be noted that, this power domain is not available for all modern processors. 

The root directory of the Intel RAPL tool within the \textit{powercap} interface is $/sys/devices/virtual/powercap/intel-rapl$. In this directory two parent power zones, namely, \textit{intel-rapl:0} and \textit{intel-rapl:1} are available, which corresponds to the CPU package domains. Inside each of the power zones, multiple sub-power zones are available, which refers to the PP0, PP1, and DRAM power domains. Several power/energy consumption and constraint attributes in each of these domains allow the user to monitor and control the respective power consumption from the admin privileged space.

\vspace{-4mm}
\section{Related Work}\label{sec:related_work}

\subsection{Performance Counter-based Detection Techniques}

Microarchitectural attacks are mostly considered as malicious activities in a system, which can be detected in real time by profiling microarchitectural components as they leave fingerprints. Since the performance monitoring unit~\cite{intel_pcm} provides a diverse set of microarchitectural event sampling, performance counters have been widely deployed in dynamic detection tools. There are several interfaces that allow users to access performance counters such as PAPI~\cite{mucci1999papi} and Intel PCM tool~\cite{intel_pcm} that enable users to select specific events to profile. Chiappetta et al.~\cite{chiappetta2016real} proposed the first study leveraging performance counters to detect the microarchitectural attacks on cryptographic implementations using Gaussian Sampling and probability density function. Zhang et al.~\cite{zhang2016cloudradar} further applied the same detection technique on cloud environment to detect anomalies on secret encryption and decryption operations. Furthermore, Dynamic Time Warping has been applied on the performance counter values to detect cache attacks. Mushtaq et al.~\cite{mushtaq2018nights} applied supervised ML models such as Linear Discriminant Analysis (LDA), Support Vector Machines (SVMs), and Linear Regression (LR) techniques to detect cache-based attacks. Briongos et al.~\cite{briongos2018cacheshield} proposed an anomaly detection technique based on Change Point Detection (CPD) to detect the rapid changes in the time series traces to identify F+F, F+R, and P+P attacks. Finally, Gulmezoglu et al.~\cite{gulmezoglu2019fortuneteller} proposed an unsupervised anomaly detection technique by leveraging advanced Recurrent Neural Networks (RNNs) to detect a wide-range of microarchitectural attacks.

\vspace{-5mm}
\subsection{Energy Consumption-based Malware Detection Techniques}

Energy consumption traces as source of anomaly detection techniques have been investigated in low-complexity systems such as 8-bit microcontrollers. To detect the anomalies caused by malware, high resolution power consumption traces are collected from simple programs. When the power consumption is larger than the pre-defined threshold value for a window period, the anomaly is detected by the detection tool~\cite{liu2016code,aguayo2011power}. Since these devices have no microarchitectural components that an attacker can target, it is not clear whether the proposed detection techniques can be implemented to detect the microarchitectural attacks in complex modern systems. Hence, several studies demonstrated that malware samples can be detected with a high accuracy with advanced analysis techniques in more complex systems such as ARM-based mobile phones through power signatures~\cite{liu2009virusmeter,caviglione2015seeing}. Moreover, Wei et al.~\cite{wei2019using} extended previous works to detect Spectre, Rowhammer, and Prime and Probe attacks on ARM devices with unsupervised ML techniques. This study is the closest work to our context as they focus on the detection of the microarchitectural attacks using power traces. However, they only demonstrate their work on the ARM-based devices to detect three types of microarchitectural attacks, which is a very small subset of the known attacks. In contrast, we profile a diverse set of microarchitectural attacks and benign applications compared to this work.

Other sensors such as electromagnetic (EM) and thermal sensors are also leveraged to detect malware and microarchitectural attacks. Zhang et al.~\cite{zhang2020leveraging} monitored hammering-correlated side-band patterns in the spectrum of the DRAM clock signal to detect Rowhammer attacks. Similarly, Vedros et al.~\cite{vedros2021limits} explored the limits of EM-based anomaly detectors in noisy environment. The EM-based detection techniques were even extended to other platforms such as medical IoT and embedded devices~\cite{sehatbakhsh2018syndrome}. Differently, Yan et al.~\cite{yan2021chip} profiled the temperature sensors located in a chip to detect anomalies in the system. This work shows that temperature differences over time can be a useful fingerprint to detect anomalies.
\vspace{-5mm}
\section{MAD-EN Dynamic Detection Tool} \label{sec:prediction_model}

\subsection{Methodology} \label{sub_sec:methodology}

\begin{figure}[t]
  \centering
  \includegraphics[scale = 0.45]{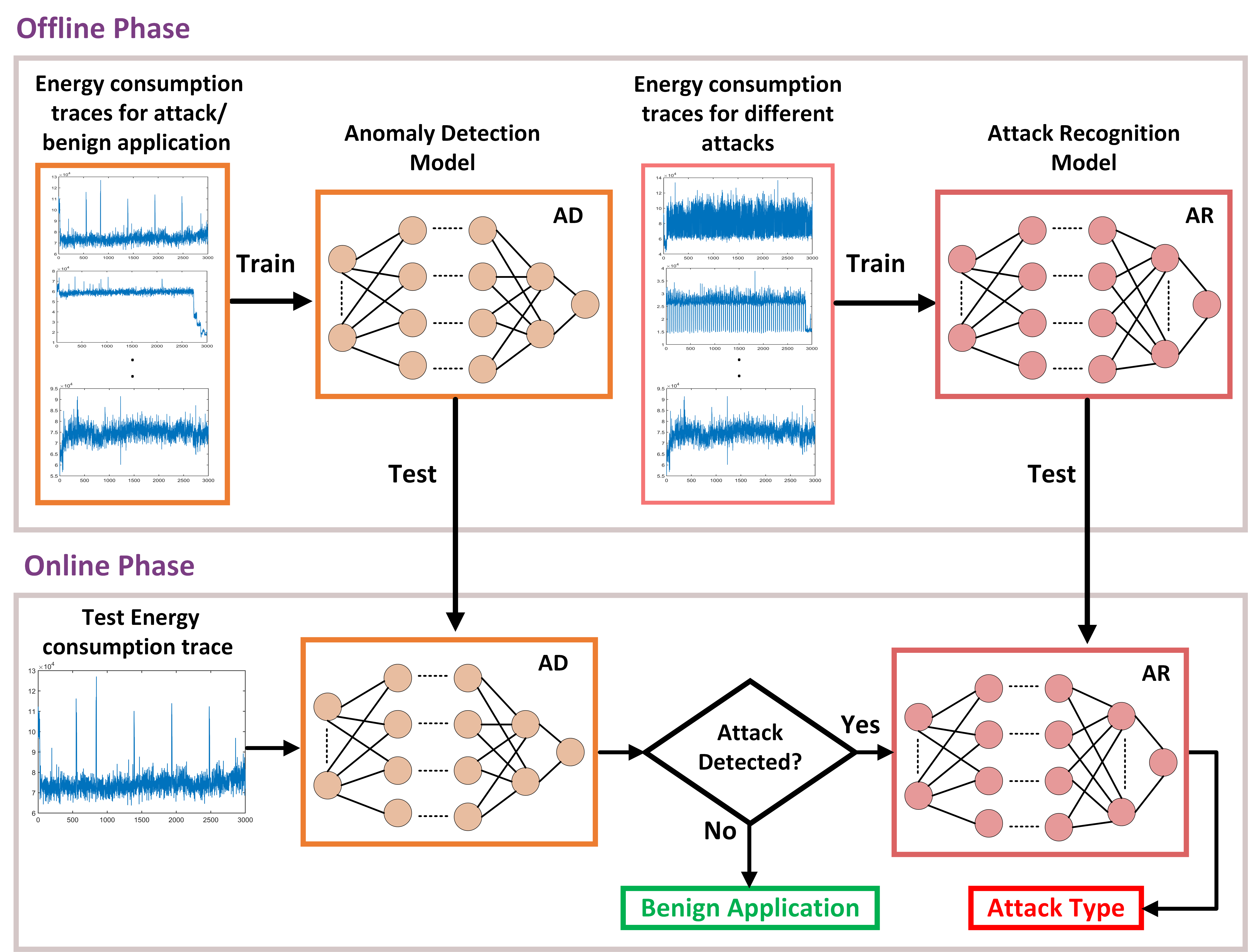}
  \caption{Methodology of the \texttt{MAD-EN} tool that comprises of two phases--offline and online phases. Two predictive models--AD and AR carried out the entire task by utilizing the system-wide energy consumption traces.}
  \label{fig:methodology}
 \end{figure}

Our purpose is to distinguish benign applications and microarchitectural attacks based on their fingerprint on CPU power consumption. For this purpose, we designed \texttt{MAD-EN} with offline and online phases as depicted in Figure~\ref{fig:methodology}. In the offline phase, a diverse set of microarchitectural attacks and benign applications is run on the test setup while the system-wide power consumption traces are collected. Next, the collected traces are utilized to train an Anomaly Detector (AD) model to be used in the online phase to detect ongoing attacks. Furthermore, an additional DL model, namely Attack Recognizer (AR), is created with solely microarchitectural attacks to classify the suspicious activity.

In the online phase, both trained models are integrated into the test device to evaluate the efficiency of the trained models in real time. Initially, a captured energy consumption trace is classified with the AD model to detect whether any attack is executing or not in the system. If the test input is predicted to be an attack execution, the model flags it as an anomaly. The AR model is implemented in a conditional manner, i.e., the same captured trace is fed to the AR model only when the AD model detects an anomaly. Therefore, if an attack is detected, AR is capable of identifying the specific attack to warn the system admin or users regarding the current threat in real time.

\vspace{-4mm}
\subsection{Data Collection} \label{sub_sec:data_collection}

As discussed previously in Section \ref{sec:background}, the Intel RAPL interface provides accumulative energy consumption data for different power domains. The micro-architectural attacks generally leverage the CPU and memory components. The attacks on uncored devices, such as GPUs, are not in the scope of our work. Therefore, the PP1 domain that includes the energy consumption of the uncored devices are not considered in the data collection. The Package domain also incorporates the overall energy consumption of both PP0 and PP1 domains, thus profiling energy consumption in the PP0 domain provides more useful information to \texttt{MAD-EN} about the ongoing microarchitectural attacks in the system. 
As a result, \texttt{MAD-EN} proceeds with the energy consumption profiling from the PP0 domain of the RAPL interface that refers to the cored devices. The $energy\_\mu J$ attribute of the PP0 domain inside the \textit{powercap} framework is utilized to collect energy consumption traces. \texttt{MAD-EN} records the energy consumption differences between two consecutive readings to observe the changes during the execution of each attack/benign application, which creates the final energy trace to be tested with AD and AR models.

In total, 15 variants from 10 different micro-architectural attacks are profiled to cover microarchitectural anomalies as the complete list is given in Appendix~\ref{appendix} Table~\ref{tab:list_of_attacks}.  
The energy consumption values are sampled through Intel RAPL interface, and 3000 samples create one trace that is used to determine an anomaly with the AD model. To prepare the train and validation datasets for the offline phase, 50 traces are collected for each microarchitectural attack. In the online phase, an anomaly test dataset is formed by launching each attack for additional 10 times to verify the viability of \texttt{MAD-EN} in a real-world system. 

The Phoronix-test-suite~\cite{phoronix_test_suite} incorporates a wide range of benchmark tests within its framework that resembles the benign workload. In this study, 25 benign applications are chosen from the Phoronix-test-suite during the energy consumption collection. Additionally, we created benign workload on commonly used applications such as Libre Office, Pycharm, Visual Studio, Website browsing (Google Chrome), and Zoom as well as streaming five videos on Youtube to enrich the set of benign applications. Thus, in total 35 benign applications are profiled for the experimentation as listed in Appendix~\ref{appendix} Table~\ref{tab:list_of_benign_apps}. The energy consumption dataset is collected during individual run-time of the applications in the same manner described previously for the attack scenario. Each application is executed 50 times to create an adequate dataset for the prediction model to be trained and validated during the offline phase. For the online phase, additional 10 measurements from 16 benign applications are recorded to form the test dataset. Moreover, for assessing the performance of the AD model with new benign applications, we have included energy traces for additional 19 processor, 13 system, and 2 disk benchmark tests from the Phoronix-test-suite (listed in Apendix~\ref{appendix} Table~\ref{tab:list_of_benign_apps}). Thus, in the online phase we have 10 measurements of energy traces from each 50 benign applications (16 tests from offline phase + 34 new tests) to enrich the test dataset.

In Algorithm \ref{alg:data_collection}, the data collection algorithm is given for both attack and benign application scenarios. The instantaneous accumulated energy consumption value of the PP0 plane at time $t$ is represented by $E1$. The next energy consumption value referred as $E2$ is sampled after a short interval, $T_i$. The difference between two consecutive readings, $\Delta E$, is one sample of the energy consumption trace in $\mu J$. In the experiments, the interval window, $T_i$, is set to $500\mu s$, and 3000 samples ($N_s$) are collected for each measurement. The number of measurements for each category, $N_M$, is 50. Hence, the overall number of traces for 15 attacks and 35 benign applications is 2500 in total. The collected dataset is used in the offline phase to train both AD and AR models.

\begin{algorithm}[t!]
\caption{Data Collection Algorithm}\label{alg:data_collection}
\tcp{$T_i$ is the interval window, and $N_s$ is the number of samples}
\tcp{$N_M$ is the number of measurements for each categories}
\tcp{$Attack/Benign\_apps$ is the attack/benign application}
\tcp{$E1$ is the accumulated energy of PP0 power plane at time $t$}
\tcp{$E2$ is the accumulated energy of PP0 power plane at time $t+T_i$}
\tcp{$\Delta E$ is the energy difference for the time window of $T_i$}
\KwInput{$T_i,N_s,N_M,url$}
\KwOutput{$\Delta E$}
\For{$i \gets 1$ to $N_M$}{
    Run $Attack/Benign\_apps$\;
    \For{$j \gets 1$ to $N_s$}{
          $E1 \gets$ Read $energy\_\mu J$\;
          sleep $T_i$ \;
          $E2 \gets$ Read $energy\_\mu J$\;
          $\Delta E \gets E2-E1$ ;
          }
    Stop $Attack/Benign\_apps$ \;
    sleep $1s$ \;
}
\end{algorithm}

\begin{figure}[t]
  \centering
  \includegraphics[scale = 0.6]{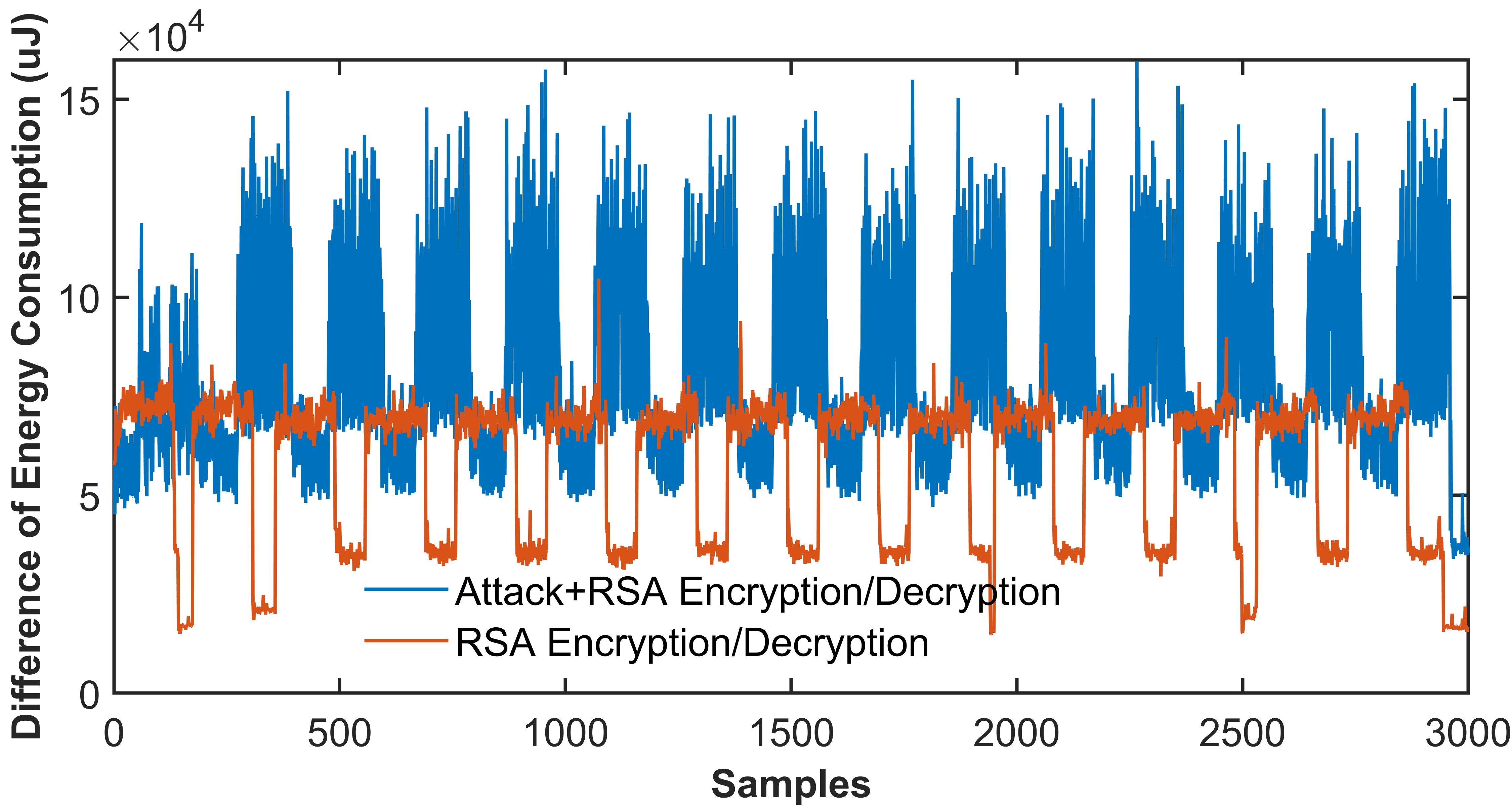}
  \caption{Two separate energy consumption traces for OpenSSL 1.1.1 RSA Encryption/Decryption (orange) and RSA+Medusa attack (blue) are illustrated in Intel(R) Core (TM) i7-10610U CPU @ 1.80GHz CPU with Ubuntu version 20.04 LTX. The Medusa v1 attack on RSA implementation has considerably higher energy consumption compared to RSA encryption/decryption. Note that, the executed attack in this scenario is Medusa attack v1.}
  \label{fig:sample_data}
 \end{figure}

Our hypothesis is that during any attack execution the system energy consumption will increase gradually compared to the targeted implementation that is running without any attack. Since one of the main targets of micro-architectural attacks is the cryptographic libraries, in which the attackers aim to steal sensitive data during the encryption or decryption, we collect two different energy consumption traces to verify our hypothesis. For this purpose, an energy consumption trace is captured while only RSA decryption is running. Similarly, a new energy consumption trace is collected during the RSA decryption while an attacker is running Medusa attack v1~\cite{moghimi2020medusa}. In Figure~\ref{fig:sample_data}, the energy consumption traces for the attack and non-attack scenarios are distinguishable since the energy consumption is doubled in presence of the Medusa attack compared to non-attack scenario. This result demonstrates that energy consumption of a system can provide sufficiently detailed information to detect ongoing microarchitectural attacks in the system.

\vspace{-5mm}
\subsection{Implementation of \texttt{MAD-EN}}
\noindent\textbf{Attack Executions:} Our purpose is to incorporate a diverse set of micro-architectural attacks including their individual variants that lead to 15 categories in total. Initially, three major cache based attacks--Flush+Flush, Flush+Reload, and Prime+Probe~\cite{github_ff} are executed to leak AES encryption keys from OpenSSL library implementation. Note that all cryptographic attacks are implemented on OpenSSL version 1.1.1.f. In the PortSmash attack~\cite{aldaya2019port}, the spy program is launched to measure the port contention delay while the TLS server is generating ECDSA signature in parallel to fully recover P-384 private key~\cite{github_portsmash}. TLBleed attack is launched for an unprivileged TLB monitoring that leads to EdDSA secret key recovery~\cite{github_tlbleed}.

The transient execution attacks have shown that the current processors are subject to different variants of Spectre attacks. In this study, we cover four variants--Spectre-PHT (v1), Spectre-BTB (v2), Spectre-RSB (v3), and Spectre-STL (v4) that are launched to leak secret information across the same user address space~\cite{canella2019systematic,github_spectre}. Moreover, one of the most recent attacks, Branch history injection (BHI)~\cite{barberis2022branch} is executed, which can be considered a new variation of Spectre-v2 attack. This attack is executed for two different modes of operations-- inter mode and intra mode. The inter mode demonstrates exploitation of unprivileged access through which arbitrary read is possible from user space to kernel space. Conversely, the intra mode abuses the privilege mode to perform random read from kernel to kernel.   

Medusa~\cite{moghimi2020medusa}, ZombieLoad~\cite{schwarz2019zombieload}, and Fallout~\cite{canella2019fallout} have distinctive characteristics, although originated from the mutation of Meltdown attack. Three variants of Medusa are executed to steal RSA keys from OpenSSL during the implementation of base64 decoding that leaves traces related to key parameters~\cite{github_medusa}. For ZombieLoad attack, we consider an attacker variant for Linux system that does not require any privilege access. This variant leaks secret information from a userspace victim application running on the same logical core~\cite{github_zombieload}. The Write Transient Forwarding (WTF) attack is executed to read arbitrary page offsets via Fallout attack. The success rate of the reads for our experimental system is 40\%, that proves its vulnerability to this attack~\cite{github_fallout}.

\noindent\textbf{AD Model:} The AD model acts as a binary classifier whose objective is to distinguish between attack executions and benign applications running within the system. In this model, all the 15 categories of attacks are clustered into a single category named anomaly and the rest of the benign applications are merged into a single group called benign. We build AD model with CNN algorithm since it outperforms other ML algorithms as further elaborated in Section~\ref{sec:evaluation}. A one-dimensional CNN model is constructed and trained with $(15+35)\times 40 = 2000$ measurements, and the remaining $(15+35)\times 10 = 500$ traces are used to validate the model. The energy consumption traces for each measurements are recorded for 30s with 3000 samples. Hence, the dimension of the input data fed into the CNN model for training is $2000 \times 3000$. The AD model comprises three convolutional layers, two max-pooling layers, and three dense layers. To overcome the overfitting issue, three dropout layers are inserted between multiple dense layers and after one of the max-pooling layer. For the three convolutional layers, the selected 1D kernel size is 3 with 64, 64, 128 filters, respectively. The activation function for all convolutional layers including the dense layers except for the last one is 'relu'. The last dense layer incorporates 'softmax' activation function in this AD model. The pooling size of the two max-pooling layers are defined as 10 to perform dimensional reduction. The AD model is tested with validation dataset during the offline phase, where 'binary cross-entropy' loss function is adopted with the 'adam' optimizer. In the online phase, the pre-trained AD model is used to carry out the detection process with new test dataset with pre-defined parameters and hyper-parameters of the model.

\noindent\textbf{AR Model:} The AR model is built to distinguish specific microarchitectural attacks in 15 categories. Similar to the AD model, the recorded energy consumption traces consist of 3000 samples for each measurement. For the offline phase, the model is trained and validated with 50 measurements from each categories, while during the online phase the model performance is tested with 10 additional measurements per categories. The CNN-based AR model incorporates four convolutional layers with a fixed kernel size of 3, followed by two max-pooling layers and four dense layers. Unlike AD model, the AR model adopts 'categorical cross-entropy' as the loss function instead of the binary. 
\vspace{-4mm}
\section{Evaluation}\label{sec:evaluation}
\vspace{-2mm}
\noindent\textbf{Experimental Setup:} The attacks and the benign applications are executed in a Dell Laptop with an x86\_64 based architecture. It accommodates Intel(R) Core (TM) i7-10610U CPU with 1.80 GHz base frequency with four physical cores. For each core, two hyperthreads are enabled. The laptop is compatible with the Intel RAPL technology that comprises four power planes--package, PP0, PP1, and dram. The Ubuntu version 20.04 LTX is the installed operating system in the laptop with a Linux kernel version of 5.11.0-46-generic. Both AD and AR models are trained and tested in a remote server of 32 cores equipped with an Nvidia GeForce RTX 3090 GPU. The server has a total of 32 cores with Intel(R) Xeon(R) Gold 5218 CPU with a base frequency of 2.30 GHz.

\begin{figure*}[t]
  \centering
  \includegraphics[scale=0.4]{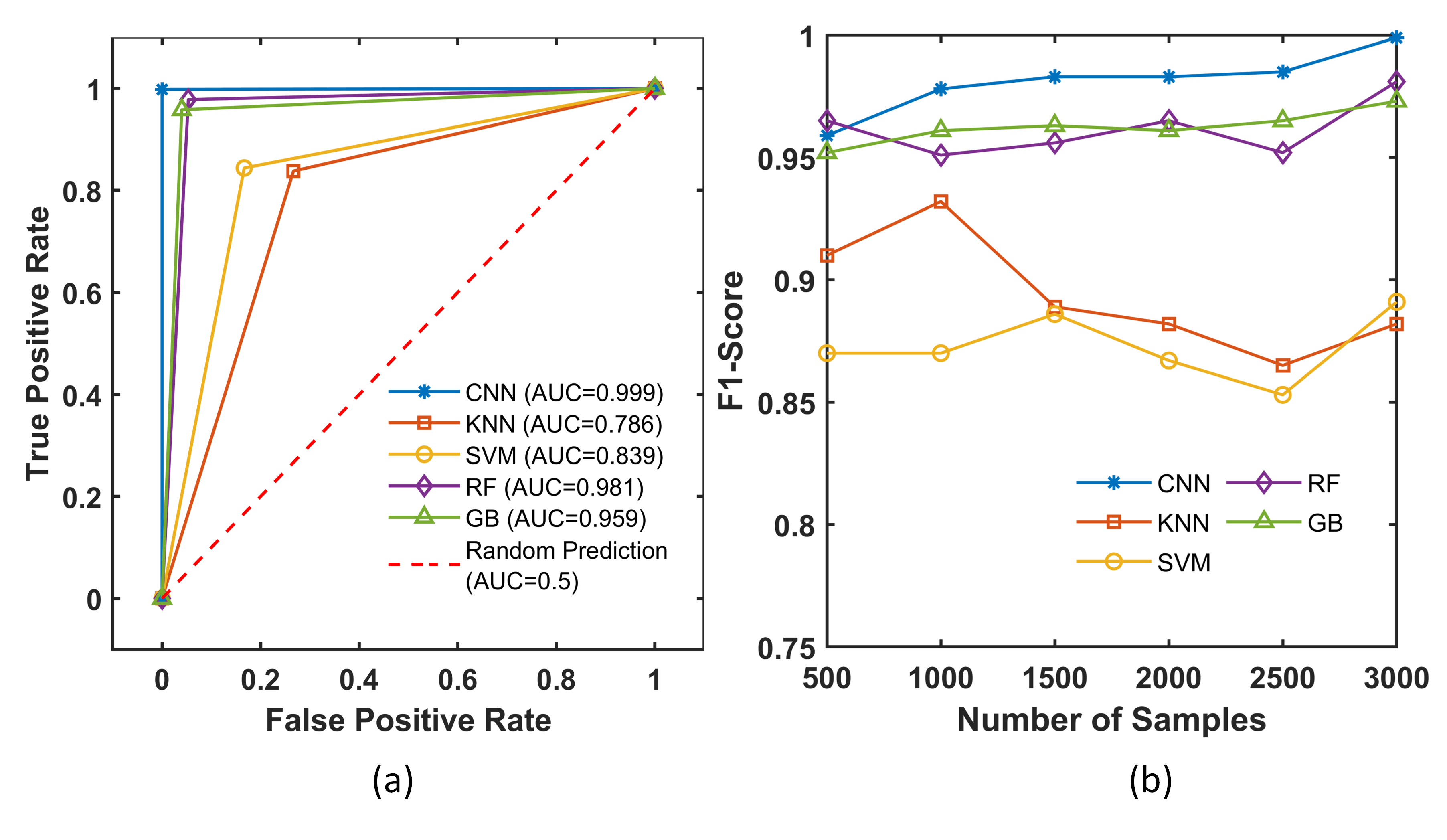}
  \caption{The performance of the AD model for CNN, KNN, SVM, RF, and GB ML and DL models is (a) illustrated with the AUC and (b) demonstrated with the F1-score based on the number of samples in the anomaly detection.}
  \label{fig:auc_f1_AD}
 \end{figure*}

\noindent\textbf{AD Model:}
The performance of the AD model is initially tested in the offline phase with the validation dataset, which consists of energy consumption traces of 350 benign measurements and 150 anomaly measurements. Once the validation accuracy does not improve further, the pre-trained model is saved to be used in the online phase. In the online phase, energy consumption traces collected from 50 different benign applications and 15 attacks are used as a test dataset, which are fed into the pre-trained AD model to evaluate its performance in a real-world scenario. The performance of CNN-based AD model is compared with four Machine Learning (ML) algorithms: Kth-nearest neighbor (KNN), Support Vector Machine (SVM), Random Forest (RF), and Gradient Boosting Tree (GB) by computing the Area under the ROC Curve (AUC), which is shown in Figure \ref{fig:auc_f1_AD}a. We show that the CNN model has the highest AUC value with a score of 0.999 compared to the other ML algorithms. Moreover, the CNN-based AD model has only 0.2\% false positive rate (FPR) and 0\% false negative rate (FNR). Both KNN and SVM models have low AUC scores, thus become ineffective for the anomaly detection. The specified parameters for the ML algorithms are listed in Appendix~\ref{appendix} Table~\ref{tab:ML_models}  
 
One of the key characteristics of micro-architectural attacks is their speed of operation. These attacks can steal useful information or secret keys within a short time frame. Hence, it is of utmost importance to detect such anomalies as fast as possible. However, faster detection mechanisms lead to the reduction of the recorded samples per measurement, considering the highest possible time resolution for energy consumption readings are maintained. The energy consumption traces for each measurement has 3000 samples, which takes around 30 seconds to collect one complete energy trace before the anomaly classification. This leads to the question, \textit{will the AD model still be effective if the data collection time frame is reduced further?}

To answer this question, the performance of the AD models is tested by decreasing the number of samples from 3000 to 500 per measurement, and a separate AD model is trained for each case. It is to be noted that, decreasing the number of samples results in faster anomaly detection as we sample the Intel RAPL framework with the highest possible resolution (500$\mu s$). Thus, the 500 samples correspond to the initial 5 seconds of the energy traces. Since the AD model is a binary classifier, F1-score of the trained models is more relevant to assess the performance of the models with different numbers of samples as given in Figure \ref{fig:auc_f1_AD}b. The F1-score decreases with the lower number of samples for the CNN-based AD model. The F1-score drops from 0.999 to 0.959 when the number of samples in a measurement is decreased to 500 samples. However, we can see that the F1-scores do not change much from 1000 to 2500 samples. Therefore, the CNN-based AD model can still be deployed in a real system with 1000 samples by compromising F1-score of 0.02 from the highest score. Interestingly, the RF-based AD model with 500 samples provides better F1-score than the CNN-based AD model. Therefore, the RF-based AD model can also be a viable option if the detection duration is the priority. While the trade-off can be made in terms of detection time and classification accuracy, the CNN-based AD model can be stated as the best one among these five ML-based AD models. 

\noindent\textbf{AR Model:} The AR model is a multi-class classifier with 15 categories that covers 10 micro-architectural attacks with their variants. The performance of the AR model is tested on the online phase with 150 energy trace measurements. Unlike the AD model, the test dataset does not contain any energy consumption traces from benign applications, as the AR model will be activated once the AD model detects an anomaly in the system. Again, the AR model is constructed with CNN, KNN, SVM, RF, and GB ML algorithms \footnote{The codes are available in GitHub: \url{https://github.com/Diptakuet/MAD-EN-Microarchitectural-Attack-Detection.git}} with different numbers of samples as listed in Table~\ref{tab:AR_accuracy}. Similar to the AD model, CNN outperforms the other four ML-based AR models with 98\% accuracy when 3000 samples are used to recognize the ongoing attacks. Interestingly, the accuracy of the KNN-based AR model does not improve even with 3000 samples. The SVM, RF, and GB-based AR models perform comparatively better only with higher number of samples, however, they have poor performance with less number of samples. Conversely, the accuracy of the CNN-based AR model does not reduce below 90\% if more than 1000 samples are used. Surprisingly, the highest accuracy of CNN-based AR model is obtained from 2500 samples instead of 3000, which is around 98.7\%.

\vspace{-5mm}
\begin{table*}[ht]
\caption{Accuracy of AR Models for 5 ML algorithms with several number of samples. The highest accuracy for each ML algorithm is highlighted.}
\centering
\label{tab:AR_accuracy}
\setlength{\tabcolsep}{16pt}
\begin{tabular}{|l|l|l|l|l|l|}
\hline
\multirow{2}{*}{Samples} & \multicolumn{5}{c|}{Test Accuracy of AR Model} \\ \cline{2-6} 
                         & CNN     & KNN     & SVM     & RF     & GB     \\ \hline
3000                     & 0.980   & 0.513   & \textbf{0.887}   & \textbf{0.867}  & \textbf{0.827}  \\ \hline
2500                     & \textbf{0.987}   & 0.43    & 0.793   & 0.760  & 0.560  \\ \hline
2000                     & 0.96    & 0.493   & 0.807   & 0.747  & 0.673  \\ \hline
1500                     & 0.967   & 0.533   & 0.820   & 0.840  & 0.673  \\ \hline
1000                     & 0.920   & \textbf{0.547}   & 0.747   & 0.731  & 0.60   \\ \hline
500                      & 0.827   & 0.52    & 0.627   & 0.645  & 0.533 \\ \hline
\end{tabular}
\end{table*}
\vspace{-5mm}

\begin{figure*}[t]
  \centering
  \includegraphics[scale = 0.65]{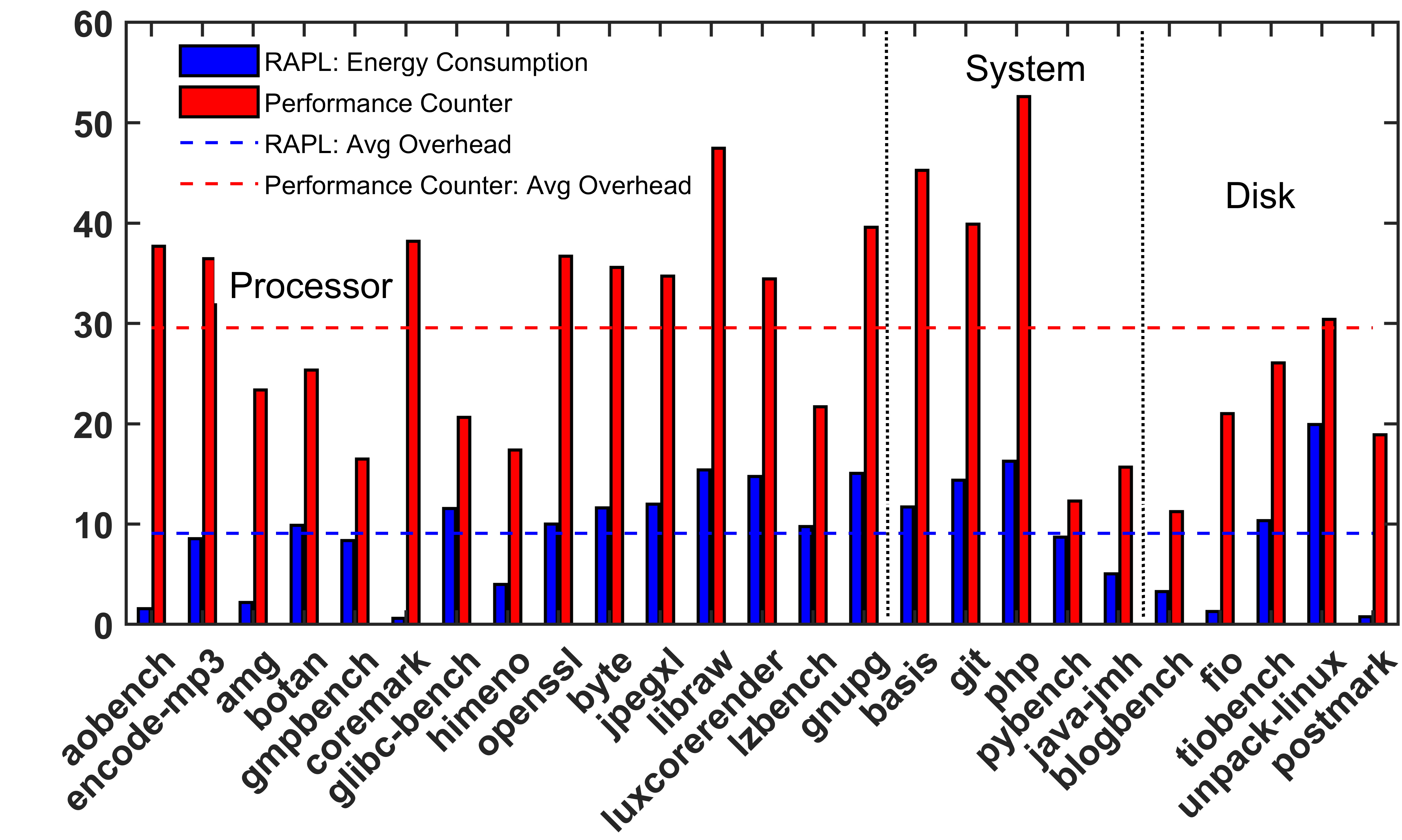}
  \caption{A comparative analysis between RAPL-energy consumption and Performance Counter features in terms of performance overhead calculated over 25 different benchmarks with similar sampling rate. MAD-EN introduces 9\% performance overhead while performance counter profiling induces 29.5\% overhead.} 
  \label{fig:perf_overhead}
\end{figure*}

\noindent\textbf{Performance Overhead:} 
The recording of energy consumption data requires hardware interaction that creates performance overhead while any execution file is running in the system. If the measurement collection creates significant performance overhead, the detection mechanism is not preferable. As many previous studies~\cite{gulmezoglu2019fortuneteller,mushtaq2018nights,mushtaq2020whisper,briongos2018cacheshield} used performance counters for the anomaly detection, a comparative analysis is performed between the performance overhead induced by the collection of energy consumption and performance counter traces. For this purpose, 25 benchmark tests are selected from the Phoronix-test-suite framework. The performance overhead for each benchmark is determined for both \texttt{MAD-EN} and performance counter-based tools with the same sampling rate of $500\mu s$ and is illustrated in Figure~\ref{fig:perf_overhead}. In the figure, the first 15 tests are from processor-based benchmarks (\textit{aobench} to \textit{gnupg}), followed by 5 system-based benchmarks (\textit{basis} to \textit{java-jmh}) and 5 disk-based benchmarks (\textit{blogbench} to \textit{postmark}).

The performance overhead comparison can be realized from Figure~\ref{fig:perf_overhead}, in which the performance counter readings create comparatively higher performance overhead than the energy consumption collection. For the energy consumption traces, the performance overhead does not exceed 20\% for any benchmarks, however, for the same benchmarks the performance overhead increases significantly for the hardware performance counter profiling. \texttt{MAD-EN} introduces 9\% performance overhead in average while performance counter profiling induces 29.5\% overhead. Based on our empirical results, proceeding with Intel RAPL tool acquires 69.3\% less performance overhead compared to the performance counter monitoring, calculated over the same 25 benchmarks.

\vspace{-4mm}
\section{Discussion}\label{sec:discussion}
\vspace{-2mm}
\noindent\textbf{Comparison with Previous Work:} In literature, most of the recent anomaly detector leverages hardware performance counter to detect micro-architectural attacks in the system~\cite{mushtaq2018nights,zhang2016cloudradar}. However, the previous studies mostly focused on differentiating cache-based attacks from non-attack scenarios. Due to limitation of adequate counters, it is quite difficult to include new attacks for the detection purposes by utilizing hardware performance counter. Gulmezoglu et al~\cite{gulmezoglu2019fortuneteller} proposed Fortuneteller in which they included some recent attacks that could successfully detect anomalies with the F1-score of 0.970. Nights-Watch is capable of detecting only the cache-based attacks with the highest accuracy of 99.54\% while considering no load scenario~\cite{mushtaq2018nights}. Our proposed \texttt{MAD-EN} model is not limited by only cache based attack detection. \texttt{MAD-EN} is capable of profiling 15 micro-architectural attacks including their variants with the highest F1-score of 0.999, which outperforms all the existing detection model. Additionally, it is imperative for any anomaly detector to have the lowest possible FPR and FNR. If the FPR becomes higher, the detection tool will end up wrongly alarming the system operator more frequently. Conversely, for higher FNR the detector will miss potential threats more often. Hence, both FPR and FNR provide fine-grained information regarding the anomaly detector. The overall FPR and FNR rate of our proposed \texttt{MAD-EN} model is 0.2\% and 0\%, respectively, which outperforms the performance of Fortuneteller~\cite{gulmezoglu2019fortuneteller}. Although the FPR of Fortuneteller and \texttt{MAD-EN} is similar, \texttt{MAD-EN} has the superiority in terms of FNR. It is to be noted that, the FPR rate of the \texttt{MAD-EN} becomes 0.2\% while testing the model with benign applications that are not included during the training phase. With known applications (included during the training phase), the FPR drops to 0\% for \texttt{MAD-EN}. 

\noindent\textbf{DRAM domain:} The Intel RAPL tool provides system-wide energy consumption from different power domains. As discussed in Section~\ref{sub_sec:data_collection}, \texttt{MAD-EN} leverages the energy consumption data recorded from the PP0 domain. Although it is expected that the DRAM domain might also be a viable source of data for profiling the attacks as these attack executions involve several memory operations, we did not observe significant changes between attack and non-attack scenarios. The energy consumption traces recorded from the DRAM domain introduce more noise to the data, thus making it ineffective for the detection. With DRAM domain based energy consumption trace, the AD model of the \texttt{MAD-EN} provides the F1-score of 0.623.

\noindent\textbf{Indistinguishable Variants:} We show that distinct variants of the attacks can also be distinguishable by MAD-EN. However, different modes of the attacks are not detectable with high confidence rate. For example, the BHI attack has two different modes of operations namely intra and inter mode. Although these two modes represent separate threat models, the energy consumption traces of these two modes of operation do not show significant differences. We believe that although the threat models are changed for these two modes, the underlying operations of these two modes remain the same, thus making them difficult to distinguish for \texttt{MAD-EN}.
\vspace{-4mm}
\section{Conclusion}\label{sec:conclusion}
\vspace{-2mm}
\texttt{MAD-EN} is the first system-wide dynamic detection tool to cover more than 10 microarchitectural attacks by utilizing energy consumption traces. \texttt{MAD-EN} can distinguish benign and malicious activities with an F1-score of 0.999, showing its applicability in real-world systems. Moreover, the ongoing attacks can be further distinguished with 98\% accuracy, which demonstrates that different attacks have distinct fingerprints on the system energy consumption. Last but not least, \texttt{MAD-EN} introduces significantly lower performance overhead compared to performance counter-based dynamic detection tools.

%
%
\bibliographystyle{splncs04}
\bibliography{Reference.bib}
\appendix
\section{Appendix} \label{appendix}

\vspace{-8mm}
\begin{table*}[h]
\tiny
\caption{List of Micro-architectural Attacks profiled in both online and offline phases}
\centering
\label{tab:list_of_attacks}
\setlength{\tabcolsep}{24pt}
\begin{tabular}{|l|l|}
\hline
1. Flush + Flush  & 5. Portsmash\\
\hline
2. Flush + Reload & 6. TLBleed\\
\hline
3. Prime + Probe  & 7. Zombieload\\
\hline
4. Spectre        & 8. Medusa\\
  -v1: Spectre-PHT      &  -v1: Cache Indexing \\
  -v2: Spectre-BTB      &  -v2: Store-to-Load Forwarding \\
  -v3: Spectre-RSB      &  -v3: Shadow REP MOV \\
  -v4: Spectre-STL      &  \\
\hline
9. Fallout & 10. Branch History Injection\\     
\hline                                                           
\end{tabular}
\end{table*}

\vspace{-12mm}
\begin{table}
\tiny
\centering
\caption{List of Benign Applications for the AD model. The 35 benign applications within the \textit{Train+Val+Test*} are utilized in the offline phase. The asterisk sign (*) refers to the 14 applications that are used as test data for the online phase. Additionally, the 34 applications within the \textit{Only Test} are the newly incorporated applications for the online phase that are not included during the training phase.}
\label{tab:list_of_benign_apps}
\begin{tabular}{|l|l|l|l|l|} 
\hline
\multicolumn{2}{|c|}{\textbf{Processor}}      & \multicolumn{1}{c|}{\textbf{System}} & \multicolumn{1}{c|}{\textbf{Disk}} & \multicolumn{1}{c|}{\textbf{Real World}}                                                                                                                                                                 \\ 
\hline
\textit{Train+Val+Test*} & \textit{Only Test} & \textit{Train+Val+Test*}             & \textit{Train+Val+Test*}           & \textit{Train+Val+Test}                                                                                                                                                                                  \\
1. aobench*              & 1. amg             & 1. git*                              & 1. blogbench                       & 1. Website stream                                                                                                                                                                                        \\
2. botan 1*              & 2. glibc-bench     & 2. php 1*                            & 2. tiobench 1                      & 2. libre office*                                                                                                                                                                                         \\
3. botan 2               & 4. gnupg           & 3. php 2                             & 3. tiobench 2                      & 3. visual studio code                                                                                                                                                                                    \\
4. botan 3               & 5. hackbench       & 4. pybench*                          & 4. unpack-linux*                   & 4. pycharm                                                                                                                                                                                               \\
5. byte*                 & 6. himeno          & 5. basis*                            & 5. fio*                            & 5. zoom                                                                                                                                                                                                  \\ 
\cline{3-4}
6. cachebench 1*         & 7. ipc-bench       & \textit{Only Test}                   & \textit{Only Test}                 & \multirow{2}{*}{\begin{tabular}[c]{@{}l@{}}6. youtube stream 1*\\\begin{tabular}{@{}l@{}}\hspace{0.5\leftmargin}--Netflix Channel\end{tabular}\end{tabular}}  \\
7. cachebench 2          & 8. libraw          & 1. financebench                      & 1. postmark                        &                                                                                                                                                                                                          \\
8. enocde-mp3*           & 9. luxcorerender   & 2. idle                              & 2. sqlite                          & \multirow{2}{*}{\begin{tabular}[c]{@{}l@{}}7. youtube stream 2 \\\begin{tabular}{@{}l@{}}\hspace{0.5\leftmargin}--HBO Channel\end{tabular}\end{tabular}}      \\
9. arrayfire 1*          & 10. lzbench        & 3. java-jmh                          &                                    &                                                                                                                                                                                                          \\
10. arrayfire 2          & 11. minion         & 4. appleseed                         &                                    & \multirow{2}{*}{\begin{tabular}[c]{@{}l@{}}8. youtube stream 3 \\\begin{tabular}{@{}l@{}}\hspace{0.5\leftmargin}--History Channel\end{tabular}\end{tabular}}  \\
11. bullet 1*            & 12. nqueens        & 5. ctx-clock                         &                                    &                                                                                                                                                                                                          \\
12. bullet 2             & 13. asmfish        & 6. hint                              &                                    & \multirow{2}{*}{\begin{tabular}[c]{@{}l@{}}9. youtube stream 4 \\\begin{tabular}{@{}l@{}}\hspace{0.5\leftmargin}--CNN Channel\end{tabular}\end{tabular}}      \\
13. bullet 3             & 14. blake2         & 7. intel-mpi                         &                                    &                                                                                                                                                                                                          \\
14. jpegxl 1*            & 15. blosc          & 8. natron                            &                                    & \multirow{2}{*}{\begin{tabular}[c]{@{}l@{}}10. youtube stream 5 \\\begin{tabular}{@{}l@{}}\hspace{0.5\leftmargin}--ESPN Channel\end{tabular}\end{tabular}}    \\
15. jpegxl 2             & 16. cloverleaf     & 9. redis                             &                                    &                                                                                                                                                                                                          \\
                         & 17. cp2k           & 10. gegl                             &                                    &                                                                                                                                                                                                          \\
                         & 18. cryptopp       & 11. sysbench                         &                                    &                                                                                                                                                                                                          \\
                         & 19. gcrypt         & 13. openscad                         &                                    &                                                                                                                                                                                                          \\
\hline
\end{tabular}
\end{table}

\vspace{-12mm}
\begin{table}[!htbp]
\tiny
\centering
\caption{The parameters used for the ML algorithms in AD and AR models (3000 samples)}
\label{tab:ML_models}
\setlength{\tabcolsep}{16pt}
\begin{tabular}{|c|l|l|} 
\hline
\multirow{2}{*}{\begin{tabular}[c]{@{}c@{}}\textbf{ML}\\\textbf{Algorithms}\end{tabular}} & \multicolumn{1}{c|}{\textbf{AD Model}}                                                         & \multicolumn{1}{c|}{\textbf{AR Model}}   \\ 
\cline{2-3}
                                                                                          & \multicolumn{1}{c|}{\textbf{Parameters}}                                                       & \multicolumn{1}{c|}{\textbf{Parameters}} \\ 
\hline
CNN                                                                                       & \begin{tabular}[c]{@{}l@{}}kernel size = 3\\stride size = 1\\padding = 'valid'\end{tabular}      & \begin{tabular}[c]{@{}l@{}}kernel size = 3\\stride size = 1\\padding = 'valid'\end{tabular} \\ \hline
KNN                                                                                       & \begin{tabular}[c]{@{}l@{}}n\_neighbors = 3\\metric = 'minkowski'\end{tabular}                 & \begin{tabular}[c]{@{}l@{}}n\_neighbors = 2\\metric = 'minkowski'\end{tabular} \\ 
\hline
SVM                                                                                       & \begin{tabular}[c]{@{}l@{}}kernel = 'rbf'\\gamma = auto\\C = 1\end{tabular}                    & \begin{tabular}[c]{@{}l@{}}kernel = 'rbf'\\gamma = auto\\C = 5\end{tabular} \\ 
\hline
RF                                                                                        & \begin{tabular}[c]{@{}l@{}}max\_depth = 14\\random\_state = 0\end{tabular}                     & \begin{tabular}[c]{@{}l@{}}max\_depth = 14\\random\_state = 0\end{tabular} \\ 
\hline
GB                                                                               & \begin{tabular}[c]{@{}l@{}}learning rate = 0.1\\max\_depth = 4\\random\_state = 0\end{tabular} & \begin{tabular}[c]{@{}l@{}}learning rate = 0.1\\max\_depth = 3\\random\_state = 0\end{tabular}\\ \hline
\end{tabular}
\end{table}
\end{document}